\documentclass{elsart}
\usepackage{natbib}
\usepackage{graphicx}
\input epsf.sty
\begin{document}
\runauthor{A.Janiuk, P. T. \. Zycki and B. Czerny}
\begin{frontmatter}
\title{X-ray reprocessing in
            Narrow-Line Seyfert 1 Galaxies:
             Ton S180
            and Ark 564}

\author[CAMK]{Agnieszka Janiuk\thanksref{kbn}}
\author[CAMK]{Piotr T. \. Zycki\thanksref{kbn}}
\author[CAMK]{Bo\.zena Czerny\thanksref{kbn}}
\address[CAMK]{N. Copernicus Astronomical Center, Bartycka 18, 00-716 Warsaw, Poland}
\thanks[kbn]{Partially supported by the Polish State Committee for 
Scientific Research, grant 2P03D018.16}

\begin{abstract}
We present the results of spectral analysis of the ASCA data for 
the Narrow-Line Seyfert~1 galaxy (NLS1) Ton~S180 and simultaneous 
ASCA and RXTE data modelling for the NLS1 Ark~564. We model both 
the primary and reflected continuum as well as the iron~K$\alpha$ line, 
the energy of which depends on the ionization state of the reprocessor. 
We show that the reprocessing matter is mildly ionized, and 
we find the soft to hard luminosity ratio to be about 2.5.
The accretion rate approximately corresponds to the Eddington limit value. 
\end{abstract}

\end{frontmatter}

\section{Introduction}

Narrow-Line Seyfert~1 galaxies are the subclass of active galaxies which 
show extremely small optical line widths, e.g. FWHM H$\beta  \le 2000$~km~s$^{-1}$, 
as well as very high luminosity and strong variability in the X-ray band.
One possible explanation of this effect is that this subclass of 
galaxies is characterized by a very high rate of accretion onto a central 
black hole of moderate mass (e.g. Pounds, Done \& Osborne 1995; 
Boller, Brandt \& Fink 1996).  In this case, the 
strong disc emission enhances the soft excess in the X-ray spectrum. In 
addition, the power-law continuum in the hard X-ray band should be steeper 
than in other Seyfert galaxies, as strong soft flux may effectively cool 
the hot disc corona. The resulting primary spectrum should therefore resemble 
that of Galactic black hole systems in the soft state. Moreover, if the 
accretion rate is high, reflection of X-rays from the ionized disc 
surface is expected. This should be accompanied with the fluorescent 
iron~K$\alpha$ line of energy 6.4--6.9~keV, depending 
on the ionization state of the reprocessor. The shape of the iron~K$\alpha$ line 
profile depends on kinematic and relativistic effects present in the inner 
regions of the accretion disc.


\section{The X-ray spectrum of Ton~S180}

The X-ray spectrum of the galaxy Ton~S180 (see Turner et~al. 1998) was modelled 
in several steps. Firstly, we fitted a 
simple power-law spectrum of photon 
spectral index $\Gamma=2.6$, with Galactic absorption fixed to $1.5\times 10^{20}$~cm$^{-2}$.
The fit was clearly not satisfactory  and the residuals show a strong spectral feature 
below 2 keV. We modelled the soft X-ray feature as a multitemperature 
disc spectrum by adding a ``disk blackbody'' model (Mitsuda et al. 1984).
However, this model for the soft continuum may not be the best one in this 
case as the spectrum shows a few distinct features due to intrinsic emission 
and absorption in the source. Therefore, we modelled the soft component with the 
optically thick Comptonization of soft photons (``comptt'' model in XSPEC),
and we included an additional absorption edge of energy $E_{\rm Edge}=0.6$~keV.
Next we included in the model the reflected continuum and the 
iron~K$\alpha$ line (\. Zycki, Done \& Smith 1997). The reprocessed 
spectrum is parametrized by the ionization parameter $\xi=L_{X}/n_{\rm e} r^{2}$ 
(Done et~al. 1992) and the reflection amplitude 
$f=\Omega/2 \pi$, where $\Omega$ is the solid angle subtended by the 
reprocessor as viewed from the X-ray source. We find that the reprocessed 
spectrum originates in a mildly ionized ($\xi \approx 200$) medium, 
which together with the assumed temperature fixed to $10^{6}$~K implies 
the iron line energy to be 6.7~keV.

\begin{table*}
\caption{Results of model fitting to the ASCA data for Ton S180. 
Model~A is a power-law spectrum with Galactic absorption, model~B has 
the additional component ``disk blackbody'' in the soft range, 
model~C also includes the reflected spectrum and iron line, while 
model~D is a sum of a power-law spectrum, the reflected component with 
iron line, the ``comptt'' model, and an absorption edge 
($kT$ is the temperature of the soft photons).  }
\label{TonS180res}
\begin{center}
\begin{tabular}{l l l l l l l}
\hline
Model  & kT & $\Gamma$ & $\Omega/2 \pi$ & $\xi$ & $E_{\rm edge}$ & $\chi^{2}/\nu$ \\
      & [keV] &        &                &       & [keV] &  (d.o.f.)    \\
\hline
A  & -- & $2.73 \pm 0.02 $ & -- & -- & -- & 1.23 (860)\\
B & $0.23 \pm 0.01$ & $2.29 \pm 0.06$ & -- & -- & -- & 1.04 (858) \\
C & $0.24 \pm 0.01$ & $2.35 \pm 0.1$ & $ 0.56^{+0.75}_{-0.35} $ & $50^{+200}_{-50} $ & -- &1.04 (947)\\
D & $0.14\pm 0.02$ & $2.56 \pm 0.01$ & $1.38 \pm 0.55$ & $150^{+350}_{-145}$ & $0.59 \pm 0.03$ & 0.99 (942)\\
\hline
\end{tabular}
\end{center}
\end{table*}

\begin{figure}
\hspace*{1in} \epsfxsize = 80 mm \epsfbox[50 180 560 660]{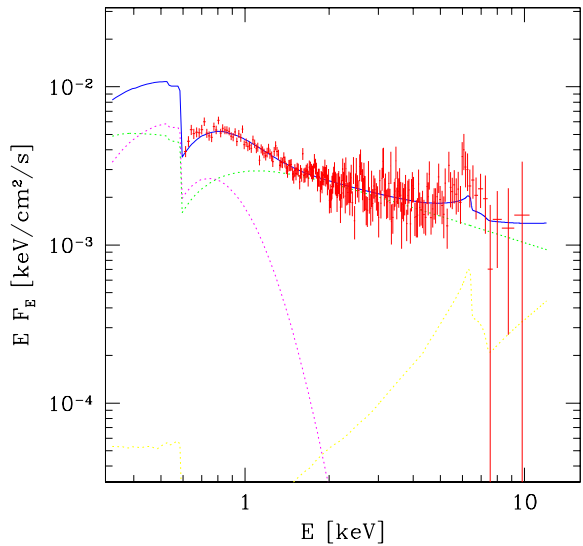}
\caption{Best-fit model to the ASCA data for Ton~S180. 
Dotted lines show the model components, the solid line shows
the resulting total spectrum, and crosses indicate the data from 
the ASCA SIS0 detector.
%
%
\label{fig:Ton}}
\end{figure}

\section{The X-ray spectrum of Ark~564}

We checked if our results are characteristic of other sources classified 
as Narrow-Line Seyfert~1 galaxies. Here we present the model for the spectrum 
of Ark~564 (Vaughan et~al. 1999). The primary X-ray continuum was modelled 
as a sum of a multicolor accretion disc and power law, 
corrected to Galactic absorption 
($N_{\rm H}$ fixed to $6.4\times 10^{20}$~cm$^{-2}$), 
and then the reflected X-ray spectrum and the iron~K$\alpha$ line were added. 
In order to obtain better constraints for the reflection parameters, we 
performed a simultaneous fit to the RXTE data. We found that the secondary  
spectrum is well described by the reflection from an ionized, flat 
disc (ionization parameter $\xi \approx 500$, reflection 
amplitude $\Omega/2 \pi\approx 1.0$), which extends almost up to the 
minimal stable orbit around the central black hole. 

\begin{table*}
\caption{Results of model fitting for the Ark~564 data. Model~A is the power-law 
spectrum with Galactic absorption, model~B also includes a disk blackbody 
component, while in model~C the reflected continuum and iron $K\alpha$ line were 
also added. Model~D has the same components as model~C but in this fit the 
additional data set from RXTE (3--20~keV) was fitted simultaneously with the ASCA data.}
\label{Ark564res}
\begin{center}
\begin{tabular}{l l l l l l l}
\hline
Model  & kT & $\Gamma$ & $\Omega/2 \pi$ & $R_{in}$  & $\xi$ & $\chi^{2}/\nu$ \\
      & [keV] &          &   & $[R_{G}]$&                   & (d.o.f.)       \\
\hline
A  & -- & $2.68 \pm 0.01$ & -- & -- & -- & 1.31 (1218)\\
B & $0.17^{+0.02}_{-0.01} $ & $2.51^{+0.03}_{-0.02}$ & -- & -- & -- & 0.99 (1216) \\
C & $0.12 \pm 0.03$ & $2.69 \pm 0.04 $ & $1.33^{+0.66}_{-0.35} $ & $16.9^{+40.0}_{-10.0} $ & $1196^{+600}_{-1000} $ & 0.96 (1217)\\
D & -- & $2.68 ^{+0.04}_{-0.05} $ & $0.96 \pm 0.36$ & $6.0^{+7.0}_{-0.0} $ & $525^{+1400}_{-370} $ & 0.91 (1029)\\
\hline
\end{tabular}
\end{center}
\end{table*}

\section{Interpretation}

The spectrum of Ton~S180 clearly shows the presence of a relativistically 
broadened iron emission line. Neglecting the effect of line broadening 
worsens the fit by $\Delta \chi^2 = 5.5$. The rest-frame line energy 
as well as the ionization parameter of the reflected spectral component 
indicate the presence of mildly ionized medium. We conclude that the 
K$\alpha$ line is produced in the ionized, upper layer of the accretion 
disc around the black hole and together with the reflected continuum resulting
from the irradiation of the plasma by hard X-rays. The reflection amplitude 
favors an isotropically illuminated disc, so that the hard X-ray emission 
may come from a flat corona. The disc inclination angle giving the best fit 
to the reflected continuum shape is in the range 30--50$^\circ$. This 
result makes the case against the explanation of observed differences between 
NLS1s and other Seyfert~1 galaxies by specific orientation to the line of sight.
The soft X-ray excess can be successfully modelled with
Comptonized black body emission and may originate in a transitionary, two-phase 
layer between the cold accretion disc and hot corona (e.g. Ro\.za\'nska et al. 1999). 
Assuming that the soft spectral component extends down to 0.01~keV, we obtain 
an absorption corrected soft luminosity of 
$L_{\rm soft}\sim 1.6 \times 10^{44}$~erg~s$^{-1}$, while the hard luminosity 
is $L_{2-10\rm keV} \sim 6.8 \times 10^{43}$~erg~s$^{-1}$. Therefore we obtain 
a soft-to-hard luminosity ratio of $\sim 2.4$. A similar method applied 
to Ark~564 gives a soft-to-hard luminosity ratio of $\sim 2.6$.
This can be explained if the system accretes with near Eddington 
luminosity, similarly to Galactic black hole binaries in their soft states.

\end{document}